\documentclass[american,aps,manuscript]{revtex4-1}
\usepackage{lmodern}

\usepackage[T1]{fontenc}
\usepackage[latin9]{inputenc}
\usepackage{units}
\usepackage{amsmath}
\usepackage{amssymb}
\usepackage{graphicx}
\usepackage{esint}

\makeatletter
\usepackage{units}

 \@ifundefined{textcolor}{}{%
   \definecolor{BLACK}{gray}{0}
   \definecolor{WHITE}{gray}{1}
   \definecolor{RED}{rgb}{1,0,0}
   \definecolor{GREEN}{rgb}{0,1,0}
   \definecolor{BLUE}{rgb}{0,0,1}
   \definecolor{CYAN}{cmyk}{1,0,0,0}
   \definecolor{MAGENTA}{cmyk}{0,1,0,0}
   \definecolor{YELLOW}{cmyk}{0,0,1,0}
 }

\usepackage{babel}

\usepackage{caption}
\newcommand{\sign}{\mathop{\mathrm{sgn}}}

\@ifundefined{showcaptionsetup}{}{%
 \PassOptionsToPackage{caption=false}{subfig}}
\usepackage{subfig}
\makeatother

\usepackage{babel}
\begin{document}

\title{Newtonian wormholes}

\author{Jos\'{e} P. S. Lemos}

\email{joselemos@ist.utl.pt}

\affiliation{Centro Multidisciplinar de Astrof\'{i}sica - CENTRA,
Departamento de F\'{i}sica, Instituto Superior T\'{e}cnico - IST,
Universidade de Lisboa - UL, Av. Rovisco Pais 1, 1049-001
Lisboa, Portugal}

\author{Paulo Luz}

\email{paulo.luz@ist.utl.pt}

\affiliation{Centro Multidisciplinar de Astrof\'{i}sica - CENTRA,
Departamento de F\'{i}sica, Instituto Superior T\'{e}cnico - IST,
Universidade de Lisboa - UL, Av. Rovisco Pais 1, 1049-001
Lisboa, Portugal}
\begin{abstract}
A wormhole solution in Newtonian gravitation, enhanced through an
equation relating the Ricci scalar to the mass density, is
presented. The wormhole inhabits a spherically symmetric curved space,
with one throat and two asymptotically flat regions. Particle dynamics
in this geometry is studied, and the three distinct dynamical radii,
namely, the geodesic, circumferential, and curvature radii, appear
naturally in the study of circular motion. Generic motion is also
analysed. A limiting case, although inconclusive, suggests the
possibility of having a Newtonian black hole in a region of finite
(nonzero) size.

\end{abstract}
\maketitle

\section{Introduction}

When many alternative theories to general relativity are being
suggested, and the gravitational field is being theoretically and
experimentally put to test, it is interesting to try an alternative to
Newton's gravitation through an unexpected but simple modification of
it. The idea is to have Newtonian gravitation, not in flat space, as
we are used to, but in curved space. This intriguing possibility has
been suggested by Abramowicz \cite{Abramowicz1}. It was noted that
Newtonian gravitation in curved space, and the corresponding Newtonian
dynamics in a circle, distinguishes three radii, namely, the geodesic
radius (which gives the distance from the center to its perimeter),
the circumferential radius (which is given by the perimeter divided by
$2\pi$), and the curvature radius (which is Frenet's radius of
curvature of a curve, in this case a circle). Using these ideas, the
perihelion advance formula for a planet in an elliptical orbit in the
gravitational field of a star in spherical space in curved Newtonian
gravitation has been computed \cite{Abramowicz1}. A further
development appeared in \cite{Abramowiczellis} where it was proposed
that besides considering Poisson's equation for Newtonian gravitation
in curved space, one could add an equation linking the
three-dimensional Ricci scalar to the matter density
\cite{ehlers,ellis}, yielding an enhanced Newtonian gravitation.  It
was argued that even having Newtonian gravitation as a limit of
general relativity the limit could yield a space with nonzero Gaussian
curvature and that observers could find that the curvature of space is
a function of the distance to the center of attraction.  In this
curved Newtonian gravitation, one can postulate a vacuum spherical
symmetric curved space solution, and then calculate the perihelion
advance of a planet in such a space, as well as the bending of light
rays \cite{Abramowiczellis}. The calculations in \cite{dadhich1} have
corroborated this idea where the field equations of Einstein were
solved by appealing to how Einstein gravity covers Newtonian gravity
when one deals with orbits of test particles.

By opening up the possibility of having Newtonian gravitation in
curved space a plethora of new situations can appear.  One can
immediately ask whether such an enhanced theory has objects that are
the analogues of the black holes, a concept that sprang from general
relativity \cite{dewitt}.  The dark stars of Mitchel and Laplace
\cite{israel} could be candidates to fill the gap.  These are stars so
compact that the escape velocity from their surface is equal or
exceeds the speed of light.  A realization of such stars was provided
by Chandrasekhar in his celebrated study of white dwarfs
\cite{ch1,ch2}. By using Newtonian gravitation and studying the full
structure of the matter in those stars, he found that diminishing the
radius from star configuration to star configuration, and finally
sending the radius to zero, there was a limiting mass, a maximum mass,
called now the Chandrasekhar mass.  Of course, in the final stages of
the sequence of configurations, there is a star, the first star that
is a dark star in the sense of Mitchel and Laplace. The other
configurations with lesser radii, including the one with $r=0$, also
belong to the class of dark stars.  However, it is known that the dark
stars of Michell and Laplace suffer severe problems in the analogy
with the general relativistic black holes.  Indeed, since in Newtonian
gravitation there is no particular status for the velocity of light,
it is difficult to argue that stars cannot be seen through light when
sufficiently compact, and even if this was the case, one could resort
to tachyonic particles that could be detected after all
\cite{penrosemichelllaplace,mcvittie}. One can try to remedy the
situation by ingenious reasoning \cite{eisen} but in the end it is
fruitless \cite{raich}, i.e., in one manner or the other one can find
ways that enable light and test particles to escape from the
Michell-Laplace dark stars (see also \cite{preti}). The true Newtonian
black holes are point-like gravitational attractors (point particles
if one prefers), as at the location of the attractor the gravitational
field is infinite and so there is no possibility of escape, one would
need an infinite acceleration and thus an infinite velocity to escape.
Indeed, the $r=0$ location of a point gravitational attractor in
Newtonian gravitation has one property that somehow fits with what is
called a singularity in general relativity: it is a singularity of the
gravitational field, the gravitational field and the tidal forces go
to infinity at $r=0$.  On the other hand it has two properties that
fit the definition of an event horizon of a black hole: (i) it is not
a singularity of space, since space is flat, and (ii) nothing can
escape from it, defining thus a horizon.  On top of that point
particles in Newtonian gravitation have in common with black holes
the fact that they have no hair, only their mass (and spin and
electric charge if that is the case) characterize the particle. So in
pure Newtonian gravitation point gravitational attractors are the
analog of the general relativistic black holes.  But now, given that
there is an enhanced Newtonian gravitation
\cite{Abramowicz1,Abramowiczellis} one can ask if there are Newtonian
black holes with a finite, non-zero radius, horizon.  Paczinski
potential \cite{pacz1,abram1}, an ad hoc potential that simulates the
effect of a horizon in the study of accretion disks in Newtonian
gravitation, is a candidate that would have to be derived from first
principles.  We have not been able by direct means to solve the
problem for this potential or for any similar one that could possess a
horizon.

But, another, simpler problem can be solved.  In particular, a curved
space in Newtonian gravitation allows the possibility of building
Newtonian wormholes, i.e., spaces not simply connected, having a
nontrivial topology that permit a passage through it. In general
relativity, wormholes have a long existence but their appearance can
be traced to the work of Morris and Thorne \cite{morristhorne}. They
were put on a firm basis in Visser's book \cite{visserbook}, and
several works have showed other special features of this system, see,
e.g.,
\cite{visserhoch,visserkardadhich,lemosetal,zaslavskiiwh,eiroa,kuhfittig,usmani-kuhfittig,balakinlemos,lemosdias,bronnikovetal,diemer,azharimousavi,dzu,but}.
The main features of a wormhole are a throat and some type of
asymptotics.  In addition, in general relativity the matter that
supports the throat violates the average null energy condition
\cite{morristhorne,visserbook}.  In particular as emphasized in
\cite{morristhorne} this means that there are always some observers
that see a negative energy density.  What can we say in enhanced
Newtonian gravitation? We will show by presenting a simple Newtonian
wormhole how these features translate into this Newtonian gravitation.

The paper is organized as follows. In Sec.~\ref{fund} we write the
fundamental equations of the enhanced Newtonian gravitation. In
Sec.~\ref{newtonworm} we construct a Newtonian wormhole, static and
spherically symmetric. We display the requisites for its construction
and then give the matter and the metric. We perform an embedding and
show the condition for having a throat, give the geometry in final
form, and study the spatial geodesics in such a space. We then analyze
the gravitational field, the gravitational potential, the mass, and
the pressure support. We also study test particle motion and dynamics
in the wormhole geometry and gravitational field. Finally, in
Sec.~\ref{newtbhs} we discuss a limiting case which suggests the
possibility of having a Newtonian black hole in a region of finite
(nonzero) size. In Sec.~\ref{conc} we conclude.

\section{The fundamental equations}
\label{fund}

Newtonian gravity was devised for an infinite flat absolute 3-dimensional
space such that Poisson's equation $\nabla^{2}\,\phi=4\pi\, G\,\rho$
holds, where $\phi$ is the gravitational potential at some spatial
point, $\nabla^{2}$ is the Laplace operator, $G$ is the gravitational
constant and $\rho$ is the mass density
of the continuum. To this equation one should
add the continuity and Euler equations for the continuum. The equation
of motion for a test particle in the gravitational field $\phi$ is
$F^{i}=ma^{i}$, where $F^{i}$ is the gravitational force given by
$F_{i}=-m\nabla_{i}\phi$, $\nabla_{i}$ is the gradient in the $i$th
spatial direction, with $i=1,2,3$, $m$ is the test particle's mass,
and $a^{i}$ its acceleration. There is, however, nothing in Newton's
theory of gravitation that forbids that Newtonian gravitation be applied
to a curved space or even a finite space. Indeed, it was recently
proposed by Abramowicz \cite{Abramowicz1} and collaborators 
\cite{Abramowiczellis}
a formulation of Newtonian gravitation in curved space.

In curved space, Poisson's equation is modified to 
\begin{equation}
g^{ij}\nabla_{i}\nabla_{j}\phi=4\pi\, G\,\rho\,,\label{curvedpois}
\end{equation}
where $g_{ij}$ is the metric for the given curved space, and the
indices $i,j$ are spatial indices running from 1 to 3. For static
systems the continuity equation is trivial, and the Euler equation
reduces to 
\begin{equation}
\nabla_{i}\, p+\rho\nabla_{i}\,\phi=0\,,\label{eulereq}
\end{equation}
where we have assumed minimal coupling, so that terms involving the
curvature of the three-space do not directly appear in the equation.
These are Newton's equations for a given imposed geometry for the
space. They are self-contained.

Now, following general theory of relativity, where the spacetime geometry
and matter are connected, it would be interesting to have, in Newton's
gravitation, the geometry of space and the matter directly related
through a new equation. One candidate for such an equation was proposed
in \cite{Abramowiczellis}, 
\begin{equation}
R=2k\rho\,,\label{connectgeomwithmatter}
\end{equation}
where $R$ is the Ricci scalar, the factor 2 appears for convenience,
and $k$ is, in this context, an arbitrary constant, that can be picked
up from observations or consistency arguments. This is an enhanced
Newtonian gravitation \cite{Abramowiczellis}. Other relations,
different from Eq.~(\ref{connectgeomwithmatter}) can be sought for.

To study the motion of a test particle with mass $m$ in such a
gravitational field $\phi$ one uses Newton's second law,
\begin{equation}
m\, a^{i}=-m\, g^{ij}\nabla_{j}\phi\,,\label{curvedeom}
\end{equation}
where $a^{i}$ is the particle's acceleration, and we have assumed
that the test particle's inertial mass $m$ is equal to the gravitational
mass.

The full enhanced theory comprises
Eqs.~\eqref{curvedpois}-\eqref{curvedeom}.

\section{Newtonian wormholes}
\label{newtonworm}

\subsection{Premises and Assumptions}

Making use of Eqs.~(\ref{curvedpois})-(\ref{connectgeomwithmatter})
we can build wormholes within the framework of Newton's theory of
gravity, and then through Eq.~(\ref{curvedeom}) one studies the
motion of test particles in such a geometry.

Our starting point is to impose a geometry that has a throat that
connects two asymptotic branches, so that there is a wormhole. We
also assume a matter density $\rho$ that is (i) finite and differentiable
everywhere, and (ii) goes to zero sufficiently fast as the distance
from the wormhole throat goes to infinity. The first condition is
justifiable in order to keep space a smooth manifold. As for the second
condition, it is imposed to guarantee that at sufficiently far distance
from the wormhole's throat we recover flat space and the usual form
for the Newton's potential. So imposing a geometry and a density compatible
with each other, one then finds the gravitational potential and the
mass of the wormhole, the pressure support, and the behavior of test
particles.

\subsection{Construction of the wormhole}

\subsubsection{Matter density and metric}

Our purpose is to construct a spherical symmetric space with geometry of a wormhole.
For this we consider its metric to be of the
form 
\begin{equation}
ds^{2}=A\left(r\right)dr^{2}+r^{2}\left(d\theta^{2}+
\sin^{2}\theta\,d\varphi^{2}\right)\,.
\end{equation}
This general form of the metric allows us to find the expression 
for the
Ricci scalar $R$ in terms of the function $A$ as 
\begin{equation}
R=\frac{2\left[\left(A\left(r\right)-1\right)A\left(r
\right)+rA'\left(r\right)\right]}{r^{2}A\left(r\right){}^{2}}\,.\label{eq:5}
\end{equation}

Found the general expression for the Ricci scalar, we can impose a
physical meaningful matter density and find 
then $A(r)$ using Eq.~\eqref{connectgeomwithmatter}. So, in order
to proceed with the construction of the Newtonian wormhole, we assume
the form of the mass density to be 
\begin{equation}
\rho\left(r\right)=\alpha e^{-\frac{r^{2}}{b^{2}}}\left(2-
\frac{b^{2}}{r^{2}}\right)\,,\label{eq:mass_density}
\end{equation}
where $\alpha$ and $b$ are two parameters with dimensions of mass
density and distance, respectively. Notice that, since we expect the
geometry of the  space to be that of a wormhole, the coordinate $r$
obeys $r>0$. Therefore, our imposition of the mass density to be
finite everywhere is verified.

Given Eqs.~\eqref{eq:5} and \eqref{eq:mass_density} we can write
Eq.~\eqref{connectgeomwithmatter} as 
\begin{equation}
\frac{2\left[\left(A\left(r\right)-1\right)A\left(r\right)+
rA'\left(r\right)\right]}{r^{2}A\left(r\right)^{2}}=2k
\alpha\frac{e^{-\frac{r^{2}}{b^{2}}}\left(2r^{2}-b^{2}
\right)}{r^{2}}\,.
\end{equation}
The solution for $A$ is then 
\begin{equation}
A\left(r\right)=\frac{r}{r+\beta r\, 
e^{-\frac{r^{2}}{b^{2}}}+C_{1}}\,,\label{eq:7}
\end{equation}
where $C_{1}$ is a constant of integration and we have defined $\beta$
as 
\begin{equation}
\beta\equiv b^{2}k\alpha\,.\label{beta}
\end{equation}
Note that $\beta$ is a pure number. Therefore, the general form of
the metric of the space, for the imposed mass density, is 
\begin{equation}
ds^{2}=\frac{r}{r+\beta\, re^{-\frac{r^{2}}{b^{2}}}+
C_{1}}\,dr^{2}+r^{2}\left(d\theta^{2}+\sin^{2}\theta\, 
d\varphi^{2}\right)\,.\label{eq:9}
\end{equation}

\subsubsection{The value of the integration constant}

Although we have found the general form of the metric of the
spherically symmetric space for the imposed matter density, we still
have to restrict the values of the parameters $k$, $\alpha$ and
$C_{1}$ to have a wormhole space. To do so, we will follow the method
adopted by Morris and Thorne \cite{morristhorne}. We use an embedding
diagram, that is, we construct in a 3-dimensional Euclidean space, a
2-dimensional surface with the same geometry of an equatorial slice,
$\theta=\pi/2$, of the spherically symmetric metric given in
Eq.~\eqref{eq:9}.

Using cylindrical coordinates $(r,z,\varphi)$ on the Euclidean embedding
space, the Euclidean metric can be written as 
\begin{equation}
ds^{2}=dr^{2}+r^{2}d\varphi^{2}+dz^{2}\,.
\end{equation}
The embedded surface will be axially symmetric and thus can be described
by a single function $z=z(r)$, such that, on that surface the line
element is 
\begin{equation}
ds^{2}=\left[1+\left(\frac{dz}{dr}\right)^{2}
\right]dr^{2}+r^{2}d\varphi^{2}\,.\label{eq:11}
\end{equation}
This line element will be the same as that of an equatorial slice
through the wormhole space, so comparing 
Eq.~\eqref{eq:9} with $\theta=\pi/2$
to Eq.~\eqref{eq:11},
identifying the coordinates $\left(r,\varphi\right)$ of the embedding
Euclidean space with the coordinates $\left(r,\varphi\right)$ of
the wormhole space, we find that 
\begin{equation}
\frac{dz}{dr}=\pm\sqrt{\frac{r}{r+\beta\, 
re^{-\frac{r^{2}}{b^{2}}}+C_{1}}-1}\,.\label{eq:12}
\end{equation}
This equation can now be used to study the proprieties of the wormhole
space. Taking the limit of Eq.~\eqref{eq:12} when $r\to+\infty$
we find that 
\begin{equation}
\lim_{r\to+\infty}\frac{dz}{dr}=0\,,
\end{equation}
for every value of the constants $C_{1}$ , $k$ and $\alpha$, concluding
that the condition of asymptotic flatness of the wormhole space is
always verified.

Next we shall analyze the throat condition, i.e., we verify that when
the coordinate $r$ goes to a minimum positive value, which we shall
set to be $b$ - defining the geometric meaning of the parameter $b$
in Eq.~\eqref{eq:mass_density} - the embedding surface has a vertical
slope, thus 
\begin{equation}
\lim_{r\to b^{+}}\frac{dz}{dr}=\pm\sqrt{\frac{1}{1+
\beta\, e^{-1}+\frac{C_{1}}{b}}-1}=\pm\infty\,.\label{eq:14}
\end{equation}
For this equation to be verified we must have 
\begin{equation}
C_{1}=-b\left(1+\frac{\beta}{e}\right)\,.\label{eq:15}
\end{equation}
Substituting this value for the constant $C_{1}$ in Eq.~\eqref{eq:14}
we find that \break $\lim_{r\to
b^{+}}\frac{dz}{dr}=\pm\sqrt{\frac{1}{\sign\left(e-
\beta\right)}}\,\infty\,.$
Thus, from the throat condition, Eq.~\eqref{eq:14}, besides finding
the value of the integration constant $C_{1}$ as given in
Eq.~\eqref{eq:15}, we also find an upper bound for the product
$\beta\equiv k\alpha\, b^{2}$, namely, $-\infty<\beta<e\,.$ This
restriction to $\beta$ is also implied from the flare out condition,
$d^{2}r/dz^{2}>0$, at the throat. Taking into account
Eq.~\eqref{eq:15} we can write Eq.~\eqref{eq:9} as
\begin{equation}
ds^{2}=\frac{1}{1-\frac{b(r)}{r}}dr^{2}+r^{2}
\left(d\theta^{2}+\sin^{2}\theta d\varphi^{2}
\right)\,,\label{eq:18}
\end{equation}
where 
\begin{equation}
b\left(r\right)=b+\frac{\beta}{e}\left(b-re^{
1-\frac{r^{2}}{b^{2}}}\right)\,,\label{eq:19}
\end{equation}
is the usually called the shape function.

To finalize the definition of the geometry of the Newtonian wormhole,
there is still one more restriction that the space must verify in
order to be a wormhole. Although the radial coordinate $r$ is ill
behaved near the wormhole's throat, the proper radial distance 
\begin{equation}
r_{*}\left(r\right)=\int_{b}^{r}\sqrt{
\frac{1}{1-\frac{b(r')}{r'}}}\, dr'\,,
\label{properrad}
\end{equation}
where the function $b\left(r\right)$ is defined by Eq.~\eqref{eq:19},
must be finite throughout the space. This implies that 
\begin{equation}
1-\frac{b(r)}{r}\geqslant0.
\end{equation}
Working out the details this implies 
\begin{equation}
-\infty<\beta\leqslant\beta_{{\rm crit}}\,,\label{eq:beta_range}
\end{equation}
where 
\begin{equation}
\beta_{{\rm crit}}\equiv\inf_{1\,\leq\, \frac{r}{b}\,<\,
\infty}\left[\frac{\frac{r}{b}-1}{e^{-1}-\frac{r}{b}\, 
e^{-\frac{r^2}{b^2}}}\right]\,,
\label{eq:beta_crit}
\end{equation}
and $\inf$ represents the infimum of the function in
parenthesis over its domain. The infimum of the right hand side of
Eq.~\eqref{eq:beta_crit} cannot be found analytically, but using
numerical methods we find that
\begin{equation}
\beta_{{\rm crit}}=2.338\,,\label{eq:beta_crit_aprox}
\end{equation}
up to the third decimal place. Comparing
Eq.~\eqref{eq:beta_crit_aprox} with the one found from the flare out
condition, namely, $\beta<e$, we conclude that the upper bound to the
product $\beta=k\alpha\, b^{2}$ is then given by
Eq.~\eqref{eq:beta_crit_aprox} itself.

The class of wormholes we are interested in here are the one that
obeys $0<\beta\leqslant\beta_{{\rm crit}}$ together with $\alpha>0$,
and so $k>0$, i.e., for this class we have wormhole spaces with
positive density and positive connection between the density and the
curvature. 
That $\beta$ can have positive values means
that it is possible to have non-exotic wormholes 
in this enhanced Newtonian gravitation. This is in contrast
to general relativistic wormholes where 
the flare out condition imposes some form 
of exoticity for the matter.

Note, however, that from the range of the parameter $\beta=k\alpha\,
b^{2}$ given in Eq.~\eqref{eq:beta_range} there are three classes that
can be considered: (i) $0<\beta\leqslant\beta_{{\rm crit}}$ -- this
class represents wormoles that can also be subdivided into (a)
wormoles with non-exotic matter, 
$\alpha>0$, and so $k>0$, and  (b) 
wormoles with exotic matter
$\alpha<0$, and so $k<0$, i.e., an exotic connection
between the curvature and the density; (ii)
$\beta=0$ -- this
class can be further subdivided into 
three subclasses: (a) if $b=0$, $k\neq0$, and $\alpha\neq0$, then the
space is trivially flat, (b) if $\alpha=0$, $b\neq0$, and $k\neq0$,
then the space is vacuum, nevertheless is curved with perhaps
interesting properties, and (c) if $k=0$, $b\neq0$, and $\alpha\neq0$,
then there is no connection between the curvature and the density and
the space can also be curved; (iii) $-\infty<\beta<0$ -- this class
represents
wormoles that can be subdivided into (a) wormoles with non-exotic matter
$\alpha>0$, and
so $k<0$, i.e., a space with positive density and negative 
exotic connection
between the density and the curvature, (b) wormoles with exotic matter
$\alpha<0$, and so $k>0$,
i.e., 
a space with negative density and positive connection between
the density and the curvature. We will exploit the interesting class
(i)(a), the wormhole spaces with non-exotic 
matter (i.e., positive density) and positive
connection between the density and the curvature.
All classes could be explored.

\subsubsection{The wormhole metric in final form in three coordinate systems}

\noindent \textit{(i) Metric in the original coordinates:}

\noindent Gathering all the results of the previous sections we can
now write the metric of the Newtonian wormhole as 
\begin{equation}
ds^{2}=\frac{1}{1-\frac{b}{r}\left(1+\frac{\beta}{e}-
\frac{\beta}{b}\, re^{-\frac{r^{2}}{b^{2}}}\right)}\, 
dr^{2}+r^{2}\left(d\theta^{2}+\sin^{2}
\theta d\varphi^{2}\right)\,.\label{eq:18xx}
\end{equation}
The Ricci scalar is given by 
\begin{equation}
R=2k\alpha e^{-\frac{r^{2}}{b^{2}}}\left(2-
\frac{b^{2}}{r^{2}}\right)\,,\label{eq:ricci1}
\end{equation}
whereas the density $\rho(r)$ is given in 
Eq.~\eqref{eq:mass_density}.

\vspace{0.3cm}

\noindent \textit{(ii) Metric in wormhole coordinates:}

\noindent Now that we have written the metric of the Newtonian
wormhole in its final form, Eq.~\eqref{eq:18xx}, we have to deal with
the problem of the coordinate singularity at $r=b$. To remove this
singularity we have to find a change of coordinates such that the
components of the metric are finite throughout the space. Considering
a new coordinate $l$ in the range $-\infty<l<\infty$, such that
$l^{2}=r^{2}-b^{2}$.  We can then rewrite the metric of the space as
\begin{equation}
ds^{2}=\frac{e\, l^{2}}{\left(b^{2}+l^{2}\right)
\left(e+\beta\, e^{-\frac{l^{2}}{b^{2}}}\right)-b
\sqrt{b^{2}+l^{2}}\left(e+\beta\right)}dl{}^{2}+
\left(l^{2}+b^{2}\right)\left(d\theta^{2}+
\sin^{2}\theta\,d\varphi^{2}\right)\,,\label{eq:final_metric_l}
\end{equation}
whose Ricci scalar is given by 
\begin{equation}
R=2k\alpha\,\frac{e^{-1-\frac{l^{2}}{b^{2}}}
\left(b^{2}+2l^{2}\right)}{b^{2}+l^{2}}.
\end{equation}
Taking the limit $l\to0$ as
we approach the wormhole's throat, we see that the coordinate
singularity is removed. Moreover, 
if the wormhole space verifies 
the condition 
Eq.~\eqref{eq:beta_range}
there are no other singularities. This form
of the metric is the most used in this work. In what follows it is
also useful to write the matter density in the new coordinates as
\begin{equation}
\rho\left(l\right)=\alpha\frac{e^{-1-
\frac{l^{2}}{b^{2}}}\left(b^{2}+2l^{2}
\right)}{b^{2}+l^{2}}.\label{eq:densitl_l}
\end{equation}

\vspace{0.3cm}

\noindent \textit{(iii) Metric in radial geodesic coordinates:}

\noindent It is also convenient to write the metric of our space in
a third coordinate system, one that uses the proper (see \cite{morristhorne})
or geodesic (see ~\cite{Abramowiczellis}) radial distance $r_{*}$
given in Eq.~(\ref{properrad}), such that the metric takes the form
\begin{equation}
ds^{2}=dr_{*}^{2}+\left[\tilde{r}\left(r_{*}
\right)\right]^{2}\left(d\theta^{2}+\sin^{2}\,
\theta\, d\varphi^{2}\right)\,,
\label{tilde}
\end{equation}
where here 
\begin{equation}
r_{*}\left(l\right)=\pm\int_{0}^{l}\sqrt{\frac{e\, 
x^{2}}{\left(b^{2}+x^{2}\right)\left(e+\beta\, 
e^{-\frac{x^{2}}{b^{2}}}\right)-b\sqrt{b^{2}+x^{2}}
\left(e+\beta\right)}}\, dx\,,\label{eq:geodesic_radius}
\end{equation}
and 
\begin{equation}
\tilde{r}^{2}\left(l\right)=b^{2}+l^{2}\,.
\label{eq:circumferential_radius}
\end{equation}
The coordinate $\tilde{r}$ in Eq.~(\ref{tilde}) is
the circumferential radius, defined such
that $2\pi\tilde{r}$ is the circumference of a circle concentric
to the wormhole's throat.

\subsubsection{Spatial Geodesics}

The spatial geodesics of the 
wormhole space should be treated with care.
For that reason we do this in the Appendix I.

\subsection{Gravitational field, gravitational potential and mass of
the wormhole}

Now that we have defined and studied the geometry of the wormhole
space we can use Eq.~\eqref{curvedpois} to find the gravitational
potential of the Newtonian wormhole. To proceed we could start by
finding the expression for the Laplace operator for this geometry and
then, using the expression for the mass density,
Eq.~\eqref{eq:densitl_l}, try to solve the differential equation for
the gravitational potential $\phi$, see Eq.~\eqref{curvedpois}. This,
however, considering the symmetries of the system, is not the most
expedited way to find the gravitational potential since, in a more
direct approach, we can use the Gauss law to find the gravitational
field and thus the gravitational potential.

We will use the metric in wormhole coordinates $(l,\theta,\phi)$, see
Eq.~\eqref{eq:final_metric_l}, and start by noticing that since the
space is spherically symmetric and the mass density is independent of
the polar and azimuthal angles, see Eq.~\eqref{eq:densitl_l}, the
gravitational field will only depend on the radial coordinate $l$,
implying that the gravitational field is constant along the surfaces
of constant $l$-coordinate.

Now, the Gauss law for gravity states that the outward gravitational
flux through any volume's boundary is proportional to the mass within
that volume. Taking into account that the space and the mass density
are also invariant under the transformation of $l\to-l$, so is the
gravitational potential $\phi$. Let us define the magnitude of the
gravitational field as
\begin{equation}
\mathcal{G}=\sqrt{g_{ij}\,\mathcal{G}^{i}\,
\mathcal{G}^{j}}\label{maggrav}
\end{equation}
where 
\begin{equation}
\mathcal{G}^{i}\equiv g^{ij}\,\phi_{,\, j}\,,\label{gravforce}
\end{equation}
is the gravitational force field, with a comma denoting a simple
derivative.  Since $\phi$ depends only on $l$, the only nonzero
component of $\mathcal{G}^{i}$ is $\mathcal{G}^{l}$. Thus,
$\mathcal{G}=\sqrt{g_{ll}\,\mathcal{G}^{l}\,\mathcal{G}^{l}}$, where
$\mathcal{G}^{l}\equiv g^{ll}\,\phi_{,\, l}$ and we see that
$\mathcal{G}$ is also invariant under the transformation $l\to-l$.
Now, the considered mass is enclosed by a surface of positive constant
$l$-coordinate, $\Sigma_{+}$, and another of negative constant
$l$-coordinate, $\Sigma_{-}$. The outward unit normal vectors to those
surfaces are given by
$n_{+}^{i}=\frac{1}{\sqrt{g_{ll}}}\,\delta_{l}^{i}$ and
$n_{-}^{i}=-\frac{1}{\sqrt{g_{ll}}}\,\delta_{l}^{i}$, where the plus
and minus signs represent the fact that if we consider the branch of
the space with positive $l$-coordinate the outward normal has the same
direction of the coordinate basis $l$-vector field and if we consider
the other branch the outward normal points in the opposite direction
of the $l$-vector field. Thus, through Eq.~\eqref{eq:final_metric_l}
the unit normals are
\begin{equation}
n_{\pm}^{i}=\pm\sqrt{\frac{\left(b^{2}+l^{2}\right)\left(e+\beta\,
e^{-\frac{l^{2}}{b^{2}}}\right)-b\sqrt{b^{2}+l^{2}}
\left(e+\beta\right)}{e\,
l^{2}}}\,\delta_{l}^{i}\,.\label{eq:normal_normalization}
\end{equation}

All is left is to define the direction of the gravitational field.
Since, for a positive mass, the gravitational force is attractive, the
outward gravitational flux through the considered volume must be
negative, which implies that the gravitational field must have
opposite direction to the normals $n_{\pm}^{i}$, therefore the
components of the gravitational field are given by
$\mathcal{G}_{+}^{l}=-\mathcal{G}^{i}\delta_{i}^{l}$ and
$\mathcal{G}_{-}^{l}=\mathcal{G}^{i}\delta_{i}^{l}$, where
$\mathcal{G}^{i}$ is given by Eq.~\eqref{gravforce}. We can now
integrate the Poisson equation, Eq.~\eqref{curvedpois}, over the
volume whose boundary are the surfaces $\Sigma_{+}$ and $\Sigma_{-}$,
such that, using the Gauss theorem
\begin{equation}
\oint_{\Sigma_{+}}g_{ij}\mathcal{G}^{i}\,
n_{+}^{j}\,\sqrt{g_{\Sigma_{+}}}d\Sigma_{+}-
\oint_{\Sigma_{-}}g_{ij}\mathcal{G}^{i}\,
n_{-}^{j}\,\sqrt{g_{\Sigma_{-}}}d\Sigma_{-}=4\pi G\,
m\left(l\right)\,,\label{eq:Gauss_flux}
\end{equation}
where $g_{\Sigma_{+}}$ and $g_{\Sigma_{-}}$ are the determinants of
the induced metric on the surfaces $\Sigma_{+}$ and $\Sigma_{-}$,
respectively, given by
$g_{\Sigma_{+}}=g_{\Sigma_{-}}=\left(l^{2}+
b^{2}\right)^{2}\sin^{2}\theta$,
and $m\left(l\right)$ is the mass within radius $l$, defined as
\begin{equation}
m(l)=\int_{0}^{2\pi}\int_{0}^{\pi}\int_{-l}^{+l}
\rho\left(x\right)\sqrt{g}\, dx\, d\theta\, d\varphi,
\end{equation}
where $g$ is the determinant of the metric of the 3-space.

Since the magnitude of the gravitational field $\mathcal{G}$,
Eq.~\eqref{maggrav}, is invariant under transformations $l\to-l$, the
gravitational flux over one of the surfaces $\Sigma_{i}$ is the same
as the gravitational flux over the other. Remembering that the
gravitational field is constant along the surfaces $\Sigma_{+}$ and
$\Sigma_{-}$ we conclude that
\begin{equation}
\mathcal{G}^{i}\left(l\right)=\frac{G\,
m\left(l\right)}{2\left(b^{2}+l^{2}\right)}
\sqrt{\frac{\left(b^{2}+l^{2}\right)\left(e+\beta\,
e^{-\frac{l^{2}}{b^{2}}}\right)-b\sqrt{b^{2}+l^{2}}
\left(e+\beta\right)}{e\,
l^{2}}}\,\delta_{l}^{i}\,,\label{eq:Grav_field}
\end{equation}
where the mass within radius $l$ is, after 
some simplification, given
by, 
\begin{equation}
m\left(l\right)=4\pi\alpha
b^{3}\int_{1}^{1+\nicefrac{l^{2}}{b^{2}}}
\frac{e^{-x}\left(2x-1\right)}{\sqrt{x\left(1+\beta
e^{-x}\right)-\left(1+\frac{\beta}{e}\right)\sqrt{x}}}\,
dx\,.\label{eq:masswithinl_integral}
\end{equation}
The integral that defines the mass within radius $l$,
Eq.~\eqref{eq:masswithinl_integral}, cannot be solved analytically
which, therefore, prevents us from finding a closed form expression
for the gravitational field. 

\begin{figure}
\centering\includegraphics[scale=0.5]{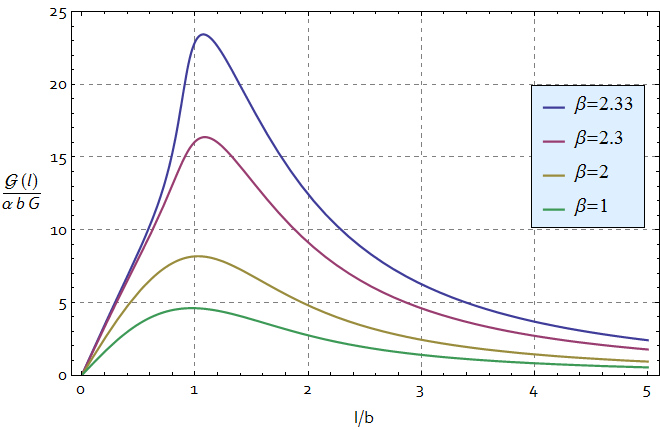}
\protect\caption{ \label{fig:grav_field}
Magnitude of the gravitational force field given by
Eqs.~\eqref{eq:Grav_field} and \eqref{maggrav} for various values of
the parameter $\beta\equiv b^{2}k\alpha$.}
\end{figure}

We then present in
Fig.~\ref{fig:grav_field} the magnitude of the gravitational field for
various values of the parameter $\beta$. In Fig.~\ref{fig:grav_field}
we consider only positive values of the radial coordinate $l$ since,
as explained in the beginning of this section, the magnitude of the
gravitational field is invariant under transformations of $l\to-l$.

\begin{figure}
\centering\includegraphics[scale=0.55]{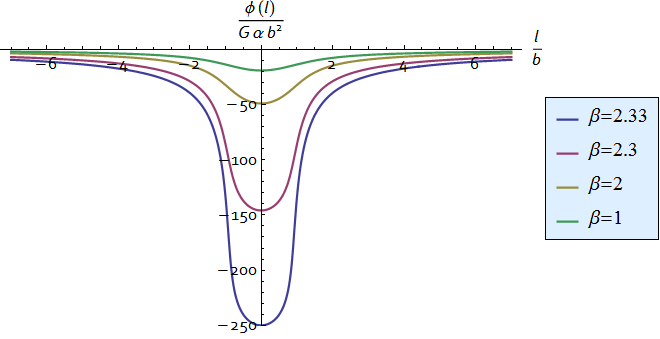}
\protect\caption{\label{fig:grav_pot}
Gravitational potential of the wormhole for the same values of the
parameter $\beta\equiv b^{2}k\alpha$ as given in
Fig.~\ref{fig:grav_field}.}
\end{figure}

Using the expression for the gravitational field,
Eq.~\eqref{eq:Grav_field}, we can find the gravitational potential
using Eq.~\eqref{gravforce}.  In Fig.~\ref{fig:grav_pot} we present
the gravitational field for various values of the parameter $\beta$,
imposing that when we move away from the wormhole's throat the
gravitational potential goes to zero.

\subsection{Pressure support on the wormhole}

\begin{figure}[b]
\centering\includegraphics[scale=0.55]{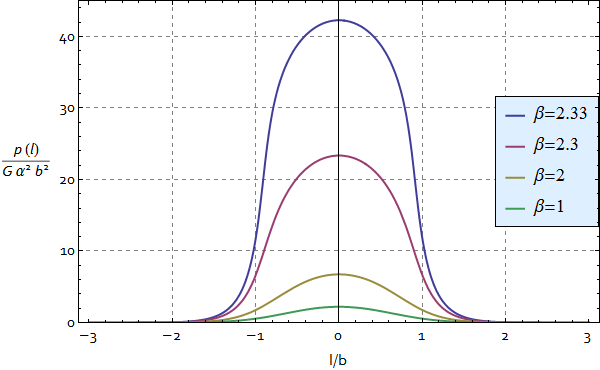} 
\protect\caption{\label{fig:pressure}
Behavior of the pressure of the fluid that permeates the Newtonian
wormhole for the same values of the parameter $\beta\equiv
b^{2}k\alpha$ as given in Fig.~\ref{fig:grav_field}.}
\end{figure}

The results found in the last section allow us to further define the
fluid that permeates the Newtonian wormhole. In the case of a static
fluid the Euler equation reduces to Eq.~\eqref{eulereq}
and in the wormhole coordinate system is given by
\begin{equation}
\frac{dp}{dl}=-\rho\frac{d\phi}{dl}\,.\label{eulereqsteadl}
\end{equation}
Since it is not possible to find a closed form expression for the
gravitational field, Eq.~\eqref{eulereqsteadl} can only be solved
numerically. The behavior of the fluid's pressure is then presented in
Fig.~\ref{fig:pressure} where it was imposed that at infinity the
pressure of the fluid is zero since the density of the fluid,
Eq.~\eqref{eq:densitl_l}, goes to zero exponentially fast as
$l\to\infty$.

Note that the pressure is positive throughout the space. This means
that matter gravitates and tends to collapse. However, the negative
curvature of space tends to have a repulsive effect on the matter,
trying to balance out the gravitational force. There is, however,
still the need of a pressure to hold the wormhole against collapse.
As is suggested by Fig.~\ref{fig:pressure}, from
Eqs.~\eqref{eq:final_metric_l}, \eqref{eq:densitl_l} and
\eqref{eq:Grav_field} we see that as the value of the parameter
$\beta$ approximates its upper bound, Eq.~\eqref{eq:beta_crit}, the
pressure of the fluid goes to infinity at the throat.

\subsection{Test
particle dynamics and
motion in the wormhole gravitational field}

\subsubsection{The equations of motion}

In the previous section we found the expression for the gravitational
field of the Newtonian wormhole, Eq.~\eqref{eq:Grav_field}. From this
gravitational field we can write the equations of motion for a test
particle subjected to the wormhole gravitational field in the wormhole
geometry. We use Eq.~\eqref{curvedeom}. We start by considering that
the particle's path is described by a curve $c$, whose components
$x^{i}$ are given by the parametric equations
$x^{i}=\left(l\left(t\right),\theta\left(t\right),\varphi
\left(t\right)\right)$.  The particle's generalized velocity
$v^{i}\equiv\dot{x}^{i}$, is given by
$v^{i}=\left(\dot{l}\left(t\right),\dot{\theta}\left(t\right),
\dot{\varphi}\left(t\right)\right)$, where a dot means derivative in
relation to time.

The left hand side of Eq.~\eqref{curvedeom} involves the acceleration
of the particle, defined as
\begin{equation}
a^{i}=v^{j}\,\nabla_{j}\, v^{i}\,=\dot{v}^{i}+
\Gamma_{jk}^{i}v^{j}v^{k}\,,\label{eq:acc2}
\end{equation}
where the $\Gamma_{jk}^{i}$ are the Christoffel symbols for the
Newtonian wormhole metric given in Eq.~\eqref{eq:final_metric_l}. The
right hand side of Eq.~\eqref{curvedeom} is the gravitational field
given in Eq.~\eqref{eq:Grav_field}. Gathering these results, we are
able to write Eq.~\eqref{curvedeom} explicitly.

First, we study the equation for the coordinate 
$\theta$. Eq.~\eqref{eq:acc2}
yields 
$\ddot{\theta}+2\frac{l}{b^{2}+l^{2}}\dot{l}\dot{\theta}-
\sin\theta\cos\theta\dot{\varphi}^{2}=0$.
%\begin{equation}
%\ddot{\theta}+2\frac{l}{b^{2}+l^{2}}\dot{l}\dot{\theta}-
%\sin\theta\cos\theta\dot{\varphi}^{2}=0\,.\label{eq:thetadotdot}
%\end{equation}
If we take the value of $\theta=\pi/2$, then this equation implies
that $\dot{\theta}=0$ is a possible solution. Checking that the higher
order derivatives are also zero, we conclude that $\theta=\pi/2$ is a
consistent solution of the equations of motion.

With this result in hand we can write the equations of motion in a
simplified form for the coordinates $l$, $\theta$, and $\varphi$, namely,
\begin{equation}
\begin{aligned}\ddot{l}+\left[\frac{b\,
e^{\frac{l^{2}}{b^{2}}}\left(2b^{2}+l^{2}\right)
\left(e+\beta\right)-2\sqrt{b^{2}+l^{2}}\left[\beta\,
l^{2}\left(1+l^{2}/b^{2}\right)+b^{2}
\left(e^{1+\frac{l^{2}}{b^{2}}}+\beta\right)
\right]}{2l\left(b^{2}+l^{2}\right)\left[b\,
e^{\frac{l^{2}}{b^{2}}}\left(e+\beta\right)-
\sqrt{b^{2}+l^{2}}\left(e^{1+\frac{l^{2}}{b^{2}}}+
\beta\right)\right]}\right]\,\,\dot{l}{}^{2}+\\
\hspace{70bp}+\left[\frac{b\sqrt{b^{2}+l^{2}}
\left(e+\beta\right)-\left(b^{2}+l^{2}\right)
\left(e+\beta\,
e^{-\frac{l^{2}}{b^{2}}}\right)}{e\,
l}\right]\,\,\dot{\varphi}{}^{2}=\\
\hspace{70bp}=-\frac{G\,
m\left(l\right)}{2\sqrt{e}\,\left(b^{2}+l^{2}\right)l}
\sqrt{\left(b^{2}+l^{2}\right)\left(e+\beta\,
e^{-\frac{l^{2}}{b^{2}}}\right)-b\sqrt{b^{2}+l^{2}}
\left(e+\beta\right)}\,,
\end{aligned}
\label{eq:ldotdot_simp}
\end{equation}
\begin{equation}
\theta=\frac{\pi}{2}\,,\quad\dot{\theta}=0\,,\label{eq:thetadotdot_simp}
\end{equation}
\begin{equation}
\ddot{\varphi}+2\,\frac{l}{b^{2}+l^{2}}\,\,\dot{l}\,
\dot{\varphi}=0\,.\label{eq:phidotdot_simp}
\end{equation}
Before trying to solve these equations of motion let us first analyze
them.

On the left hand side of Eq.~\eqref{eq:ldotdot_simp}, we have two
terms that depend on the first derivative of one of the coordinates,
one in $\dot{\varphi}^{2}$ and the other in $\dot{l}^{2}$. The former
is the usual centripetal force that also arises in classical mechanics
in flat space. However the latter term has no flat space counterpart.
This term arises indeed from the curvature of space along the radial
direction. The curvature, then, implies that a force similar to a
centripetal force arises when the particle's motion has a radial
component.  Taking the limit $l\to\pm\infty$, and recalling that the
wormhole is asymptotically flat, we find that the term in
$\dot{l}^{2}$ vanishes, i.e., in flat space it is nonexistent. Still
in Eq.~\eqref{eq:ldotdot_simp} one finds that the particle suffers an
usual centripetal acceleration $a_{{\rm c}}$ given by
\begin{equation}
a_{{\rm c}}=\left[\frac{b\sqrt{b^{2}+l^{2}}
\left(e+\beta\right)-\left(b^{2}+l^{2}\right)\left(e+\beta\, 
e^{-\frac{l^{2}}{b^{2}}}\right)}{e\, l}\right]\,
\dot{\varphi}^{2}\,.\label{acccentrep1}
\end{equation}

We also see that Eq.~\eqref{eq:phidotdot_simp} can be readily
integrated, obtaining
\begin{equation}
h=\left(l^{2}+b^{2}\right)\dot{\varphi}\,,
\end{equation}
where $h$ is a constant of integration that can be interpreted as
the angular momentum per unit mass of the particle. Thus, angular
momentum is conserved as one expects for a central force.

Once the equations of motion of a test particle are found, we have now
to solve them. We will study two cases, pure circular motion and a
general motion.

\subsubsection{Circular motion}

\begin{figure}
\centering\includegraphics[scale=0.5]{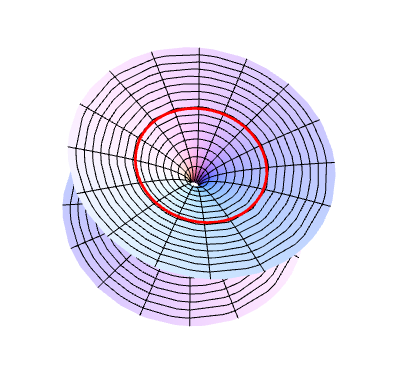}
\protect\caption{\label{fig:circular_motion} 
Circular motion of a particle in the Newtonian wormhole for some
initial radial position $l_{0}$ and initial angular velocity
$\dot{\varphi}$ given by Eq.~\eqref{eq:circular_phi}.  The wormhole
parameters are $b$ and $\alpha$.}
\end{figure}

Let us first consider the problem of circular motion of a test
particle in the Newtonian wormhole background. In this case the radial
coordinate $l$ does not change throughout the particle's motion,
therefore, $l=l_{0}$ say, $\dot{l}=0$, and $\ddot{l}=0$. From this,
Eq.~\eqref{eq:ldotdot_simp} reduces to
\begin{equation}
\ddot{\varphi}=0\,,
\end{equation}
which in turn has a first integral $\dot{\varphi}=\text{\text{constant}}$.
With the help of Eq.~\eqref{eq:ldotdot_simp} we can determine this
constant and obtain 
\begin{equation}
\dot{\varphi}=\sqrt{\frac{\sqrt{e}\, G\, m
\left(l_{0}\right)}{2\left(b^{2}+l_{0}^{2}\right)
\sqrt{\left(b^{2}+l_{0}^{2}\right)\left(e+\beta\, 
e^{-\frac{l_{0}^{2}}{b^{2}}}\right)-b
\sqrt{b^{2}+l_{0}^{2}}\left(e+\beta\right)}}}\,,
\label{eq:circular_phi}
\end{equation}
where it is implicit that the particle can rotate in both senses.
Equation \eqref{eq:circular_phi} relates the radial position $l_{0}$
of the particle and the azimuthal velocity $\dot{\varphi}$ that the
particle must have in order to describe a circular orbit around the
wormhole's throat. In Fig.~\ref{fig:circular_motion} we present
an example of such circular motion. 

This case of circular motion can also be used to check the consistency
of our treatment against the analysis presented by Abramowicz et al
\cite{Abramowicz1,Abramowiczellis}.  Thereby, our purpose is to find
the centripetal acceleration using the following expression presented
in \cite{Abramowiczellis}, i.e.,
\begin{equation}
a^{i}=-\frac{v^{2}}{\mathcal{R}}n_{\pm}^{i}\,,
\label{eq:Abramo_centripetal}
\end{equation}
where $v=\sqrt{\left(l^{2}+b^{2}\right)}\dot{\varphi}$ is the modulus
of the linear velocity of the particle in circular motion,
$n_{\pm}^{i}$ is the outside pointing unit normal to the circle given
in Eq.~\eqref{eq:normal_normalization}, and $\mathcal{R}$ is the the
curvature radius of the circle described by the particle's motion
given by \cite{Abramowiczellis}
\begin{equation}
\mathcal{R}=\tilde{r}\,\left(\frac{dr_{*}}{d
\tilde{r}}\right)\,.\label{xpto}
\end{equation}
Using Eqs.~\eqref{eq:geodesic_radius} and
\eqref{eq:circumferential_radius} for the geodesic radius $r_{*}$, and
the circumferential radius $\tilde{r}$, respectively, we find that in
this case the curvature radius is given by
\begin{equation}
\mathcal{R}=\sqrt{\frac{\left(b^{2}+l^{2}
\right)^{2}e}{\left(b^{2}+l^{2}\right)\left(e+\beta\, 
e^{-\frac{l^{2}}{b^{2}}}\right)-b\sqrt{b^{2}+l^{2}}
\left(e+\beta\right)}}\,.\label{eq:curvature_radius}
\end{equation}
Gathering all the intermediate results, and putting 
the centripetal
acceleration $a_{c}$ as $a_{{\rm c}}=a^{l}$, 
we can write Eq.~\eqref{eq:Abramo_centripetal}
explicitly has 
\begin{equation}
a_{{\rm c}}=\left[\frac{b\sqrt{b^{2}+l^{2}}
\left(e+\beta\right)-\left(b^{2}+l^{2}
\right)\left(e+\beta\, e^{-\frac{l^{2}}{b^{2}}}
\right)}{e\, l}\right]\,\dot{\varphi}^{2}\,,
\end{equation}
which is exactly the expression that we found for the centripetal
acceleration that appears in Eq.~\eqref{acccentrep1}.

\subsubsection{General motion}

\begin{figure}[ht]
\subfloat[]{\centering\includegraphics[scale=0.5]
{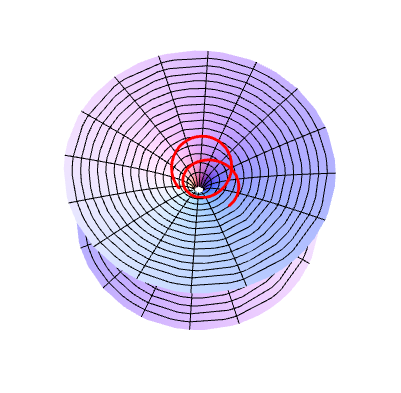}}
\subfloat[]{\centering \includegraphics[scale=0.5]
{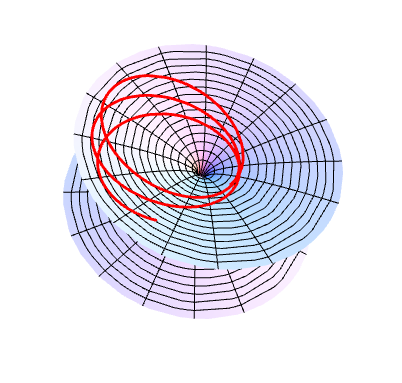}

}

\protect\caption{ \label{fig:general_motion} 
Panel (a): Particle's motion in the wormhole background for a
relatively small initial angular velocity $\dot{\varphi}$.
Panel (b): Particle's motion in the
wormhole background for a relatively large angular velocity
$\dot{\varphi}$. In all figures the particle's initial radial position
is taken to be at some $l_{0}$, and the initial radial velocity
$\dot{l}_{0}$ is zero. The wormhole parameters are $b$ and $\alpha$. }
\end{figure}

Aside the case of circular motion which can be treated analytically,
we can use numerical methods to solve the equations
\eqref{eq:ldotdot_simp}-\eqref{eq:phidotdot_simp}  to study a more
generic motion of a particle on the Newtonian wormhole given its
initial position and velocity. We present in
Fig.~\ref{fig:general_motion} some graphical solutions of
Eqs.~\eqref{eq:ldotdot_simp}-\eqref{eq:phidotdot_simp} for general
motions.

\vfill

%\subsection
\section{The limiting case: Are there Newtonian black holes?}
\label{newtbhs}

Figs.~\ref{fig:grav_field} and \ref{fig:grav_pot} suggest that as the
parameter $\beta$ approaches the upper bound given by
Eq.~\eqref{eq:beta_crit} the value of the
gravitational field $\mathcal{G}^{i}\left(l\right)$ and the
gravitational potential
itself have a somewhat weird behavior. Indeed, analyzing
Eqs.~\eqref{eq:Grav_field} and \eqref{eq:masswithinl_integral}, 
one finds that, for some value $l_{{\rm h}}$ of the
radial coordinate $l$, the 
gravitational field $\mathcal{G}^{i}\left(l\right)$
goes to infinity when the value
of the parameter $\beta$ approaches the value $\beta_{{\rm crit}}$,
given by Eq.~\eqref{eq:beta_crit}.

This is a very surprising behavior since the system somehow develops a
Newtonian event horizon, in the sense that for $\beta=\beta_{{\rm
crit}}$ every test particle that crosses the sphere defined by
$l=l_{{\rm h}}$ can only return back from that inner sphere to the
outside if it has infinite acceleration, and thus an infinite
velocity. Hence, an observer beyond that sphere can never affect an
outside observer. It characterizes a Newtonian black hole.

However, note also that as the value of the parameter $\beta$
approximates its upper bound, Eq.~\eqref{eq:beta_crit}, the fluid
pressure goes to infinity at the throat, see also
Fig.~\ref{fig:pressure}.  Thus, the black hole cannot be sustained
with this type of matter at such an extreme situation. Presumably at
such $\beta_{{\rm crit}}$, the matter collapses leaving behind two
disjoint regions in each side of the original space with a singularity
at the center where the density and the space curvature blow up.  This
final singularity formed from the collapse of the original matter
could also be considered a Newtonian black hole in this enhanced
Newtonian gravitation, since only test particles with infinite
velocities could escape from it.

\section{Conclusions}
\label{conc}
 
We have presented a Newtonian wormhole in Newtonian gravitation in
curved space, enhanced with relation between the curvature and the
matter density. The wormhole is spherically symmetric having two
connected asymptotic flat regions. The wormhole's mass density is
positive throughout the space.  This is distinct from wormholes in
general relativity where the flare out condition imposes some form of
exoticity for the matter. The wormhole is hold against collapse by
pressure.  Test particle motion in the wormhole gravitational field
have shown that the three distinct dynamical radii, namely, the
geodesic, circumferential, and curvature radii, appear naturally in
the study of circular motion.  A limiting case, when the parameter
$\beta$ satisfies $\beta=\beta_{\rm crit}$, suggests the possibility
of having a Newtonian black hole in a region of finite size.

%\newpage{}

\vskip 7cm
\section*{Appendix I: Spatial Geodesics}

Here we display the spatial geodesics of the wormhole
space.

\subsection{Spatial geodesic equations}

To complete the analysis of the Newtonian wormhole geometry, we 
find the expressions for the geodesics in this space. To start we
consider a spatial curve $c$ on the Newtonian wormhole parametrized
by a parameter $\lambda$. A point on $c$ is given by the parametric
equation 
$c^{i}\left(\lambda\right)=\left(l\left(\lambda\right),
\theta\left(\lambda\right),\varphi\left(\lambda\right)\right)$
and the tangent vector to the curve at a given point by 
$c^{i}\,'\left(\lambda\right)=\left(l'\left(\lambda\right),
\theta'\left(\lambda\right),\varphi'\left(\lambda\right)\right)$,
where a dash means derivative with respect to the affine parameter
$\lambda$. The curve $c^{i}$ is said to be a geodesic if it parallel
transports its tangent vector along itself, i.e., its tangent vector
must be such that the equation
\begin{equation}
c^{i}\,'\nabla_{i}\, c\,^{j}\,'=0\,,
\end{equation}
is verified, and where 
$\nabla$ is the covariant derivative.
Before we write the geodesic equations explicitly for
this space, we can simplify the calculations
by noticing that due to the spherical
symmetry of the space we can always choose a coordinate system in
such a way that the polar angle, 
represented by the coordinate $\theta$, is constant
along the curve and equal to $\pi/2$. 
In other words, since the $2$-sphere,
$S^{2}$, is a submanifold of the space and it is a well known
fact
that part of great circles are geodesics in $S^{2}$, then any geodesic
on the space, under isometries, is part of a great circle. Thus,
the geodesic equations for the wormhole space are
\begin{equation}
\left\{ \begin{aligned} & l''+\left[\frac{b\, 
e^{\frac{l^{2}}{b^{2}}}\left(2b^{2}+l^{2}\right)\left(e+
\beta\right)-2\sqrt{b^{2}+l^{2}}\left[\beta\, l^{2}
\left(1+l^{2}/b^{2}\right)+b^{2}\left(e^{1+
\frac{l^{2}}{b^{2}}}+\beta\right)\right]}{2l\left(b^{2}+
l^{2}\right)\left[b\, e^{\frac{l^{2}}{b^{2}}}\left(e
+\beta\right)-\sqrt{b^{2}+l^{2}}\left(e^{1+
\frac{l^{2}}{b^{2}}}+\beta\right)\right]}\right]\,\, 
l'^{2}+\\
& \hspace{4.75cm}+\left[\frac{b\sqrt{b^{2}+l^{2}}
\left(e+\beta\right)-\left(b^{2}+l^{2}\right)
\left(e+\beta\, e^{-\frac{l^{2}}{b^{2}}}\right)}{
e\, l}\right]\,\,\varphi'^{2}=0\,;\\
& \varphi''+2\frac{l}{b^{2}+l^{2}}\, l'\varphi'=0\,;\\
& \theta'=0\,,\\
& \theta=\pi/2\,.
\end{aligned}
\right.\label{eq:geodesics}
\end{equation}
The geodesic equations in their general form are quite complicated
to solve and find an analytical solution. We can however
consider simpler cases than the general one.

\subsection{Spatial azimuthal geodesics}

Let us first consider the case of purely azimuthal geodesics, i.e.,
consider geodesics of the form
$c^{i}\left(\lambda\right)=\left(l_{0},
\pi/2,\varphi\left(\lambda\right)\right)$.
In this case, the geodesic equations simplify to $\varphi''=0$. If we
take the initial and final position of the geodesic to be,
respectively,
$c^{i}\left(0\right)=\left(l_{0},\pi/2,\varphi_{0}\right)$ and
$c^{i}\left(\lambda_{f}\right)=\left(l_{0},\pi/2,\varphi_{f}\right)$
we conclude that
\begin{equation}
\varphi\left(\lambda\right)=\frac{\varphi_{f}-
\varphi_{0}}{\lambda_{f}}\,\lambda+\varphi_{0}\,.
\end{equation}

\subsection{Spatial radial geodesics}

Let us now consider the case
of purely radial geodesics. In this case the geodesics are of the
form $c^{i}\left(\lambda\right)=
\left(l\left(\lambda\right),\pi/2,\varphi_{0}\right)$,
such that the geodesic equations simplify to 
\begin{equation}
l''+\left[\frac{b\,
e^{\frac{l^{2}}{b^{2}}}\left(2b^{2}+l^{2}\right)
\left(e+\beta\right)-2\sqrt{b^{2}+l^{2}}\left[\beta\,
l^{2}\left(1+l^{2}/b^{2}\right)+b^{2}\left(e^{1
+\frac{l^{2}}{b^{2}}}+\beta\right)\right]}{2l
\left(b^{2}+l^{2}\right)\left[b\,
e^{\frac{l^{2}}{b^{2}}}\left(e+\beta\right)-
\sqrt{b^{2}+l^{2}}\left(e^{1+\frac{l^{2}}{b^{2}}}+
\beta\right)\right]}\right]\,\,
l'^{2}=0\,.\label{eq:radial_geodesics}
\end{equation}
This equation is still too complicated and it does not seem to be
possible to find an analytical solution to it. We can, however,
consider a further simplification, namely, we
consider that the initial
and final radial coordinate, $l_0$
and $l_f$, respectively, 
are very close to the wormhole's
throat. This allow us to Taylor expand the expression in parenthesis
in Eq.~\eqref{eq:radial_geodesics} such that
\begin{equation}
l''+\left[\frac{l\left(e-3\beta\right)}{4b^{2}
\left(\beta-e\right)}\right]\,\, l'^{2}=0\,.
\end{equation}
This expression can be analytically integrated. If we then consider
the initial and final points to be
$c^{i}\left(0\right)=\left(l_{0},\pi/2,\varphi_{0}\right)$ and
$c^{i}\left(\lambda_{f}\right)=\left(l_{f},\pi/2,\varphi_{0}\right)$,
respectively, we find that
\begin{equation}
\begin{aligned}l\left(\lambda\right) & =
-i\sqrt{\frac{8b^{2}\left(\beta-e\right)}{e-
3\beta}}\text{erf}^{-1}\left(i\,\frac{
\lambda_{f}-\lambda}{\lambda_{f}}\text{erfi}
\left(\sqrt{\frac{e-3\beta}{8b^{2}\left(\beta-e\right)}}\, 
l_{0}\right)+\right.\\
 & \left.\hspace{160bp}+\frac{i\,\lambda}{
\lambda_{f}}\text{erfi}\left(\sqrt{\frac{e-
3\beta}{8b^{2}\left(\beta-e\right)}}\, l_{f}\right)\right)
\end{aligned}
\label{eq:radialsol_geodesics}
\end{equation}
where $\text{erf}^{-1}$ represents the inverse Gauss error function
and $\text{erfi}$ the imaginary Gauss error function. Equation
\eqref{eq:radialsol_geodesics} gives the expression for the radial
coordinate of purely radial geodesics when the two considered points
are very close to the wormhole's throat.

\subsection{Spatial general geodesics}

The simplifications adopted in the previous subsections greatly
simplified the geodesic equations and allowed us to find closed form
solutions for those simplified cases. Let us, however, tackle the
general problem of studying geodesics of the form
$c^{i}\left(\lambda\right)=\left(l\left(\lambda\right),
\pi/2,\varphi\left(\lambda\right)\right)$,
with initial and final points
$c^{i}\left(0\right)=\left(l_{0},\pi/2,\varphi_{0}\right)$ and
$c^{i}\left(\lambda_{f}\right)=\left(l_{f},\pi/2,\varphi_{f}\right)$,
respectively. Notice that we imposed the angular coordinate
$\theta=\pi/2$.  This reflects the argument that we presented in the
beginning of this section, i.e.,
given any two points on the spherically
symmetric Newtonian wormhole we can always find an isometry such that
the geodesic that connects those two points is part of a great circle.

To start, we introduce a new equation: 
\begin{equation}
l'^{2}\left(\frac{e\, l^{2}}{\left(b^{2}+l^{2}\right)
\left(e+\beta\,
e^{-\frac{l^{2}}{b^{2}}}\right)-b\sqrt{b^{2}+l^{2}}
\left(e+\beta\right)}\right)+\varphi'^{2}
\left(b^{2}+l^{2}\right)=1\,.\label{eq:55}
\end{equation}
This equation represents the restriction that the norm of the tangent
vector to the curve $c^{i}$ is constant, which we normalized and set
to $1$. Next, we integrate the azimuthal equation in
Eq.~\eqref{eq:geodesics}, obtaining
\begin{equation}
\varphi'=\frac{C}{b^{2}+l^{2}}\,,\label{eq:56}
\end{equation}
where $C$ is a constant of integration. Substituting Eq.~\eqref{eq:56}
in Eq.~\eqref{eq:55} we find that 
\begin{equation}
l'=\sqrt{\left[\frac{e+\beta\, e^{-\frac{l^{2}}{b^{2}}}}{e\,
l^{2}}-\frac{b\left(e+\beta\right)}{e\,
l^{2}\sqrt{b^{2}+l^{2}}}\right]\left(l^{2}+b^{2}-
C^{2}\right)}\,.\label{eq:57}
\end{equation}
Dividing Eq.~\eqref{eq:56} by Eq.~\eqref{eq:57} we finally find
that 
\begin{equation}
\frac{d\varphi}{dl}=C\sqrt{\frac{e\,
l^{2}}{\left[\left(b^{2}+l^{2}\right)^{2}\left(e+\beta\,
e^{-\frac{l^{2}}{b^{2}}}\right)-b\sqrt{\left(b^{2}+
l^{2}\right)^{3}}\left(e+\beta\right)\right]
\left(l^{2}+b^{2}-C^{2}\right)}}\,,
\end{equation}
or 
\begin{equation}
\varphi\left(l_{f}\right)={\displaystyle
C\int_{l_{i}}^{l_{f}}\sqrt{\frac{e\,
l^{2}}{\left[\left(b^{2}+l^{2}\right)^{2}\left(e+\beta\,
e^{-\frac{l^{2}}{b^{2}}}\right)-b\sqrt{\left(b^{2}+
l^{2}\right)^{3}}\left(e+\beta\right)\right]
\left(l^{2}+b^{2}-C^{2}\right)}}}\,\,
dl+\varphi_{0}\,,\label{eq:generalsol_geodesics}
\end{equation}
which can be numerical integrated, provided the value of the constant
$C$ is given, such that the geodesic's path is described by
$c(l)=\left(l,\pi/2,\varphi(l)\right)$.

We can further define the constant of integration $C$. Let's consider
the unit vector
\begin{equation}
X_{\varphi}^{i}=\frac{1}{\sqrt{b^{2}+l^{2}}}\delta_{\varphi}^{i}\,,
\end{equation}
defined has the tangent unit vector to curves with constant $l$ and
$\theta$ coordinates.\textbf{ }Taking the dot product of
$X_{\varphi}^{i}$ and the tangent vector of the geodesic $c^{i}\,'$,
we have, using Eq.~\eqref{eq:final_metric_l}, that 
\begin{equation}
X_{\varphi}^{i}\, c_{i}\,'=\sqrt{\left(b^{2}+
l^{2}\right)}\varphi'=\frac{C}{\sqrt{b^{2}+
l^{2}}}\,,\label{eq:61}
\end{equation}
where in the last step we used Eq.~\eqref{eq:56}. By other hand,
defining the smaller angle between $X_{\varphi}^{i}$ and $c^{i}\,'$
as $\xi$, the definition of the dot product implies that 
\begin{equation}
X_{\varphi}^{i}\, c_{i}\,'=\cos\left(\xi\right)\,,\label{eq:62}
\end{equation}
where we used the fact that the vectors $X_{\varphi}^{i}$ and $c^{i}\,'$
are unitary vectors. Comparing Eqs.~\eqref{eq:61} and \eqref{eq:62}
we find that the integration constant $C$ is related to the angle
$\xi$ by: 
\begin{equation}
C=\sqrt{b^{2}+l^{2}}\cos\left(\xi\right)\,.\label{eq:63}
\end{equation}

Incidentally, Eqs.~\eqref{eq:generalsol_geodesics} and \eqref{eq:63}
allow us to further describe the purely radial geodesics, studied
in the previous subsection where a closed form solution for such geodesics
was only possible if the beginning and ending points were very close
to the wormhole's throat. If we consider that the angle $\xi=\pi/2$
in Eq.~\eqref{eq:63}, using Eq.~\eqref{eq:generalsol_geodesics}
we find that the coordinate $\varphi\left(l\right)$ is constant along
the geodesic. This is exactly as we expected since for 
$\xi=\pi/2$ the geodesic $c^{i}$ is orthogonal to the geodesics
with constant $l$-coordinate, i,e,, $c^{i}$ is a radial geodesic.
This result allow us to describe all the radial geodesics on the 
wormhole space,
such that, $c^{i}\left(l\right)=\left(l,\pi/2,\varphi_{0}\right)$,
supplanting Eq.~\eqref{eq:radialsol_geodesics} which was valid only
when the initial and final points were very close to the wormhole's
throat.

\newpage{}

\end{document}